\documentclass{emulateapj}
\usepackage{epsfig}
                                                                                 
\shorttitle{Radiative Cooling and Intracluster Entropy}
\shortauthors{McCarthy et al.}
\begin{document} 

\title{The Effects of Radiative Cooling on the Entropy Distribution of 
Intracluster Gas}

\author{Ian G. McCarthy, Mark A. Fardal, and Arif Babul}

\affil{Department of Physics \& Astronomy, University of Victoria, Victoria, BC, 
V8P 1A1, Canada;\\ 
imccarth@uvic.ca, fardal@uvic.ca, babul@uvic.ca}

\begin{abstract}
High resolution X-ray observations indicate that the entropy profiles in the 
central regions of some massive cooling flow clusters are well approximated 
by powerlaws.  McCarthy and coworkers recently accounted for this trend with 
an analytic model that includes the detailed effects of radiative cooling.  
Interestingly, these authors found that cooling (and subsequent inflow of the 
gas) {\it naturally} establishes approximate steady-state powerlaw entropy 
profiles in the cores of clusters.  However, the origin of this behavior and 
its dependence on initial conditions have yet to be elucidated.  In the present 
study, we explain this trend in the context of the self-similar cooling wave 
model developed previously by Bertschinger (1989).  It is shown that the 
logarithmic 
slope of the entropy profile in the cores of relaxed cooling flow clusters is 
given by a simple analytic function that depends only on the logarithmic 
slopes of the local gravitational potential and the cooling function.  We 
discuss a number of potentially interesting uses of the above result, 
including: (1) a way of measuring the shapes of gravitational potentials of 
cooling flow clusters (which may, for example, be compared against the standard 
hydrostatic equilibrium method); (2) a simple method for constructing 
realistic analytic cluster models that include the effects of radiative 
cooling; and (3) a test of the reliability of cooling routines implemented in 
analytic models and hydrodynamic simulations.

\end{abstract}


\section{Introduction}

The presence of strong positive temperature gradients (e.g., Allen et al.\ 2001; 
De Grandi \& Molendi 2002; Piffaretti et al.\ 2004; Vikhlinin et al.\ 2004) and 
the lack of an obvious entropy floor (e.g., Pratt \& Arnaud 2002; Mushotzky et 
al. 2003; Piffaretti et al.\ 2004) in the cores of some clusters have 
stimulated interest in theoretical models of the intracluster medium (ICM) 
that include the effects of radiative cooling.  Such models have been shown to 
be in broad agreement with the current suite of X-ray observations of 
clusters, particularly if some form of feedback/heating is also incorporated 
into the models (e.g., Tozzi \& Norman 2001; Voit et al.\ 2002; Oh \& Benson 
2003; Borgani et al.\ 2004; McCarthy et al.\ 2004, hereafter M04).

\footnotetext[1]{We note that such ``preheating'' does not preclude subsequent 
heating events following cluster formation, such as AGN feedback initiated by the 
accretion of cold gas at the cluster center.} 

\footnotetext[2]{
However, the mild amount of heating is sufficient
to delay the onset of catastrophic cooling for up to several Gyr and, as
a result, the predicted global cold gas fractions remain consistent with
observationally established limits.}

Interestingly, by assembling published {\it Chandra} and {\it XMM-Newton} 
X-ray data from the literature, M04 uncovered a population of relaxed clusters 
(e.g., A2029, PKS0745, Hydra A) with central entropy profiles (where we define 
the ``entropy'', $S$, as $kT n_e^{-2/3}$) that are well approximated by 
powerlaws (see also the recent study of Piffaretti et al.\ 2004).  M04 
were able to account for these clusters (as well as other clusters that 
exhibit large entropy cores) with a simple analytic model that 
includes the detailed effects of cooling and that assumes the ICM was 
initially heated prior to cooling.  In particular, M04 found that if the ICM 
in a given cluster is initially heated$^1$ by only a mild amount, cooling {\it 
naturally} establishes approximate powerlaw entropy profiles near the cluster 
core$^2$.  It would be interesting to understand from a physical perspective what 
is driving this trend in the observed and model clusters.

Unfortunately, a rigorous calculation of the effects of radiative cooling 
and the subsequent (quasi-hydrostatic) inflow of intracluster gas can only 
formally be obtained by numerically solving the time-dependent hydrodynamic 
equations (as approximately done, e.g., in M04).  This makes it a challenge to 
understand physically why the above trends are established by cooling.  
Despite this inconvenience, there are, however, analytic tools at our disposal 
that can help to establish a physical picture.  For example, Bertschinger 
(1989) (hereafter, B89) used a self-similarity analysis to derive the behavior 
of cooling flows in clusters.  One of the interesting results derived by 
Bertschinger is that the logarithmic slopes of the gas density and 
temperature profiles in the limit $r \ll r_{cool}$ (where $r_{cool}$ is the 
radius at which the cooling time of the gas equals the age of the cluster) 
depend only on the shapes of the cluster gravitational potential and of the 
cooling function, $\Lambda(T)$.  The behavior of the gas density and 
temperature profiles can be used to infer how radiative cooling influences the 
distribution of intracluster entropy, which is perhaps a more fundamental 
quantity since convection will strive to prevent the establishment of a 
rising entropy profile towards the cluster center (see, e.g., Voit et al.\ 
2002).  Thus, the study of B89 is a good starting point for our investigation 
of the effects of cooling and inflow on intracluster entropy.

In the present study, we briefly review the self-similar analysis of B89, 
including the basic assumptions made in that study and their validity, and use 
his results to derive how radiative cooling modifies the entropy profiles of 
clusters.  We compare the self-similar solution to the results of the 1-D 
cooling model of M04 for clusters with powerlaw dark matter profiles and that 
cool via thermal bremsstrahlung and, indeed, find extremely good agreement in 
the limit $r \ll r_{cool}$.  We further demonstrate that the self-similar 
solution should provide an accurate description for cooling (via both 
bremsstrahlung {\it and} line emission) in more realistic dark matter halos.  
Finally, we discuss a number of interesting uses of this result, including 
measuring the shapes of cluster gravitational potentials (that, e.g., could 
be compared against the usual hydrostatic equilibrium method), testing the 
reliability of numerical cooling routines in analytic models and hydro 
simulations, and as a simple way of setting up initial conditions for cluster 
models with radiative cooling.

\section{Cooling Flows and Self-Similarity: The Bertschinger Solution}

As noted above, in general, the time-dependent hydrodynamic equations must be 
solved numerically in order to obtain a detailed picture of the effects of 
radiative cooling on cluster gas.  However, when the time dependence is due 
to a single physical process that can be characterized by a unique scale 
length [in this case, the cooling radius$^3$, $r_{cool}(t)$], similarity 
solutions can provide useful physical insight.  Adopting this approach, B89 
derived the general behavior of cooling flows in clusters of varying 
gravitational potentials (and with varying cooling functions as well).  We are 
especially interested in the properties of his solution in the limit $r \ll 
r_{cool}$, since they are independent of the initial conditions of the 
gas. (Outside the cooling radius the initial conditions are of 
crucial importance since radiative cooling hasn't had enough time to 
significantly modify the gas there.)  Before examining the solution itself, 
let us first review the basic assumptions made in B89 and their validity.

\footnotetext[3]{The cooling radius grows larger with time.  Unfortunately, 
the self-similar solution in B89 is expressed in terms of the {\it initial} 
cooling radius.  Thus, it might be expected that comparison to observations, which 
are used to infer the {\it present} cooling radius, is somewhat ambiguous.  
However, as highlighted by B89, this leads to only a small error since, over the 
course of a cluster's life, the cooling radius grows only by a small amount.}   

Since self-similar solutions can be characterized by only a single scale 
length, some simplifying assumptions are required in order to obtain a solution 
for the properties of cooling flows.  In particular, Bertschinger assumed that 
the gravitational potential and the cooling function could be characterized by 
simple powerlaws that remained fixed as a function of time.  Of course, in 
reality, neither of these assumptions are strictly valid.  High resolution 
numerical simulations indicate that the dark matter density profiles of 
clusters (and halos of other masses as well) have a characteristic scale 
length (the so-called scale radius, $r_s$), where the index of the powerlaw 
profile changes from relatively shallow (between $\sim -1$ and $-1.5$; e.g., 
Navarro, Frenk, \& White (NFW) 1997; Moore et al.\ 1999) to 
relatively steep ($\sim -3$).  The cooling function is not scale-free 
either, owing primarily to line emission.  Thus, one is justified in 
questioning the physical relevance of a model that invokes these 
assumptions. (However, physical relevance may not be of major concern if 
one is simply testing the reliability of numerical cooling methods.)
We examine in \S 3.2 whether relaxing these assumptions significantly affects 
the shapes of the resulting entropy profiles in the cores of {\it massive} 
clusters.

Other assumptions made in the analysis of B89 include spherical symmetry, 
subsonic inflow, and single-phase cooling.  Spherical symmetry is expected to 
be approximately valid, at least in an average sense for a reasonably large 
relaxed cluster sample.  Likewise, subsonic flow should hold for the majority 
of the gas within $r_{cool}$, except possibly near the very center where the 
gas may become transsonic.  However, it has yet to be determined whether or 
not multi-phase cooling is important in clusters.  In the absence of 
significant non-gravitational heating, the intracluster gas {\it is} 
thermally unstable.  However, because the gas flows into the cluster center 
essentially as fast as it cools, we expect multi-phase cooling to be 
important only near the very center (see B89).  Recent spatially-resolved {\it 
Chandra} and {\it XMM-Newton} spectra of the central regions cooling flow 
clusters have confirmed that, probably with the exception of the very central 
radial bin, single-phase models provide at least as good a fit as multi-phase 
models (e.g., David et al.\ 2001; Matsushita et al.\ 2002).  For the present 
study, we assume single-phase cooling.

Finally, it is implicitly assumed that there are no significant sources of 
non-gravitational heating (such as AGN feedback, thermal conduction, and 
turbulent mixing) present in the ICM.  In \S 4, we give a brief discussion of 
the potential impact of such heating.

Implementing the above assumptions, B89 renormalized the hydrodynamic 
equations by removing any time dependence arising through $r_{cool}(t)$.  A 
self-similar solution is obtained if one neglects the acceleration terms 
(which is valid since the inflow of gas is highly subsonic) in the 
renormalized hydro equations.  It is straightforward to derive the limiting 
behavior of the gas density and temperature profiles in the limit 
$r \ll r_{cool}$ under these conditions (see eqns. 2.30 of B89):

\begin{equation}
\frac{d\log{\rho}}{d\log{r}} = \frac{-3 + (2 - \alpha)(1 - \beta)}{2}
\ \ , \ \
\frac{d\log{T}}{d\log{r}}  =  2 - \alpha
\end{equation}

\noindent where we have assumed that dark matter dominates the gravitational 
potential and that $\rho_{dm} \propto r^{-\alpha}$ and $\Lambda(T) \propto 
T^{\beta}$.  Note that the definitions of $\alpha$ and $\beta$ differ from the 
definitions of these symbols in B89.

In \S 1, we defined the ``entropy'', $S$, in terms of gas density and 
temperature.  The above equations can, therefore, be used to yield the 
logarithmic slope of the entropy profile within $r_{cool}$:

\begin{equation}
\gamma \equiv \frac{d\log{S}}{d\log{r}} = \biggl[1 - \frac{1}{3}(1 - \beta) 
\biggr](2 - \alpha) + 1
\end{equation}

Thus, for $\beta = 1/2$ (i.e., cooling dominated by thermal bremsstrahlung),

\begin{equation}
\gamma = \frac{5}{6}(2 -\alpha) + 1
\end{equation}

\noindent which yields $\gamma = 1$ for a singular isothermal sphere 
($\alpha = 2$), $\gamma \approx 1.4$ for a Moore et al.\ profile in the limit 
$r \ll r_s$ ($\alpha = 1.5$), and $\gamma \approx 1.8$ for a NFW profile 
in the limit $r \ll r_s$ ($\alpha = 1$).  Note, however, that the value 
of $\gamma$ in equation (2) depends only weakly on the shape of the cooling 
function (that is, for reasonable values of $\beta$ ranging from $-1/2$ to 
$1/2$) for $\alpha \gtrsim 1$.

Below, we compare this simple analytic result with the 1-D cooling model of 
M04.  

\section{Comparison with M04}

\subsection{Powerlaw clusters}

M04 developed a simple radiative cooling code which, when applied to 
the entropy injection model clusters of Babul et al.\ (2002), successfully and 
simultaneously reproduces the luminosity-temperature and luminosity-mass 
relations (including their associated intrinsic scatter) and yields detailed 
fits to the entropy, surface brightness, and temperature profiles of 
clusters as inferred from recent high resolution X-ray observations.  As 
alluded to in \S 1, these authors found that radiative cooling established a 
powerlaw entropy profile in cores of their model clusters that experienced 
only mild preheating (that is, for those clusters that were initially injected 
with $\lesssim 300$ keV cm$^2$, i.e., the entropy cooling threshold for massive 
clusters).  It is interesting 
to see whether or not this trend can be explained by the self-similar solution 
of B89.

In order to test this, we use the model of M04 (see \S 2.2 of M04 
for a detailed discussion of the model) to track the effects of cooling for a 
set of clusters with arbitrary initial conditions (recall that the solution of 
B89 within the cooling radius does not depend on initial gas conditions).  For 
specificity, however, we show results for clusters that have a total dark 
matter mass of $10^{15} M_{\odot}$, a total gas mass of $\approx 
1.5 \times 10^{14} M_{\odot}$, and a maximum radius, $r_{halo}$, of $2.06$ 
Mpc.  The cluster gas is initially assumed to be in hydrostatic equilibrium 
within a dark matter halo that is characterized by a density profile 
$\rho_{dm} \propto r^{-\alpha}$ (the normalization being set by the mass and 
radius given above).  The cluster gas is then allowed to evolve via radiative 
cooling and inflow until a stable entropy profile is achieved.  To calculate 
the cooling rate, we assume a cooling function that scales as $\Lambda(T) 
\propto T^{1/2}$ with a normalization that is set by matching a zero 
metallicity Raymond-Smith plasma at a temperature of $T \approx 10^{8}$ K.  
Finally, as in M04, we remove any gas that is able to cool below a 
threshold temperature of $10^5$ K from the calculation and assume that its 
dynamical effects on the cluster gravitational potential are negligible.  
Clearly, if a significant amount of gas is able to completely cool this 
assumption will be violated.  However, in this case, we expect these effects 
will be relevant only for the central tens of kpc and will have only a minor 
effect on the overall entropy distribution within the cooling radius.

\begin{figure}
\centering
\includegraphics[width=8.4cm]{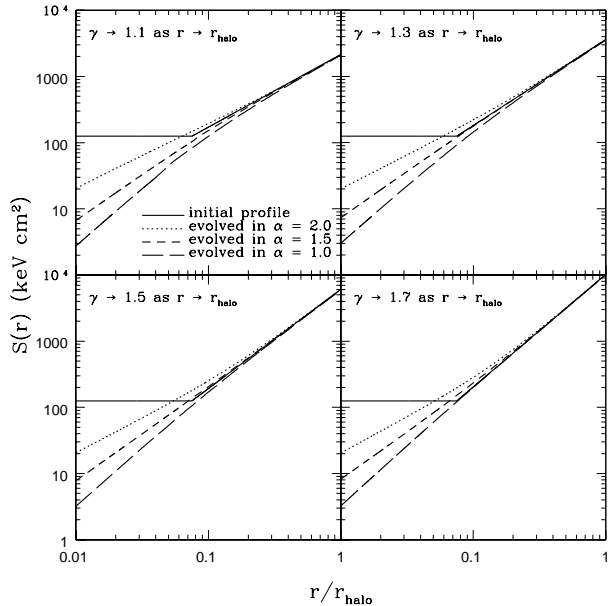}
\caption{
The effects of cooling and inflow on the entropy distribution of clusters.
Solid lines represent the initial entropy profiles.  The dotted, short
dashed, and long dashed lines represent the resulting steady-state entropy
profiles when the clusters are evolved in dark matter halos that have density
profiles characterized by powerlaw indices of $\alpha = 2.0$, $1.5$, and $1.0$,
respectively.  The various panels show the resulting profiles for
different initial entropy distributions.  Figure 2 presents a comparison of the
central logarithmic entropy slopes to the Bertschinger solution.
}
\end{figure}
\begin{figure}
\centering
\includegraphics[width=8.4cm]{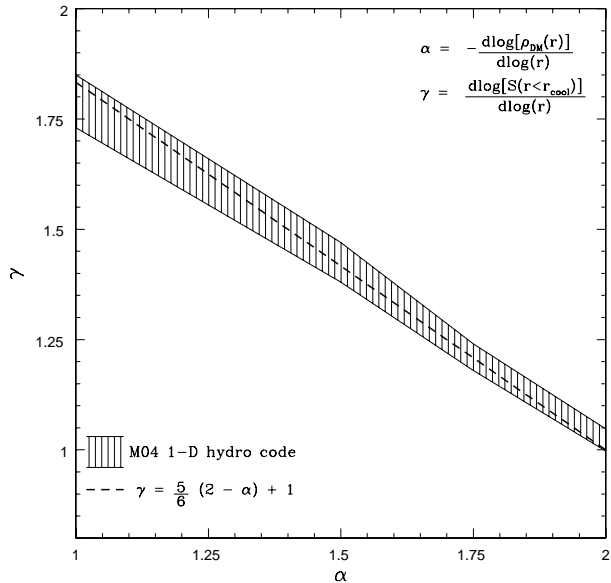}
\caption{
Comparison of the self-similar solution of B89 with the results of M04's
cooling model.  The shaded region reflects the uncertainty in the best
fit powerlaw indices of the entropy profiles (within $r_{cool}$) shown in
Fig. 1.
}
\end{figure}
\begin{figure*}
\centering
\includegraphics[width=13.0cm]{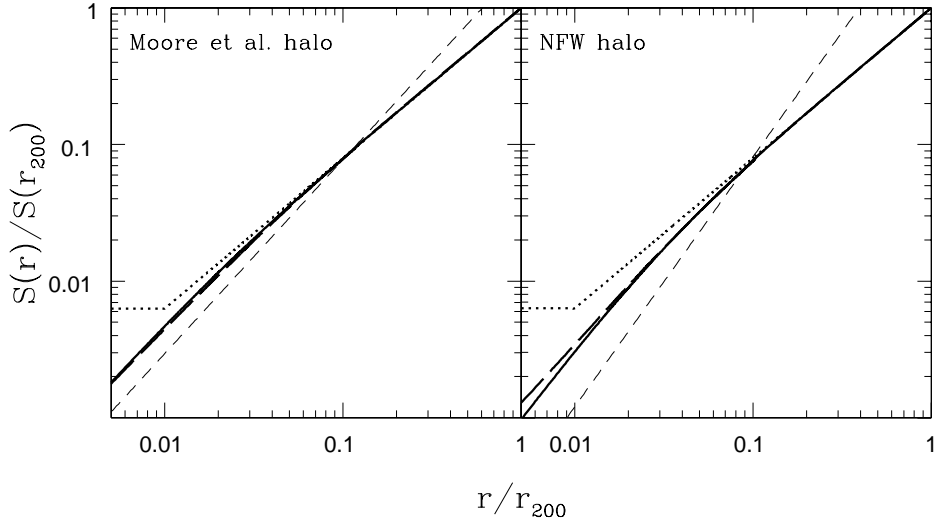}
\caption{
The effects of line emission and realistic dark matter profiles on the
steady-state entropy profile.  {\it Left:} A cluster with a Moore et al.\ dark
matter halo.  {\it Right:} A cluster with a NFW dark matter halo.  In both
panels the dotted line represents the initial entropy distribution, the
thick solid lines represent the final entropy distribution assuming cooling
with a pure thermal bremsstrahlung cooling function, and the thick dashed lines
represent the final entropy distribution assuming cooling with a Raymond-Smith
plasma model with $Z = 0.3 Z_{\odot}$.  The thin dashed lines indicate the
slope predicted by B89's self-similar model assuming $\beta = 1/2$ and $\alpha
= 1.5$ (left panel) and $\alpha = 1.0$ (right panel).  They have been
arbitrarily normalized to cross the thick lines at the cooling radius.
}
\end{figure*}

To show how the effects of cooling are linked to the underlying gravitational 
potential, we present Figure 1.  Focusing first on the top left-hand panel, 
we start with a cluster that is initially characterized by entropy profile that 
contains a core and a logarithmic slope of $\gamma = 1.1$ outside the core.  
This is the initial slope adopted by Babul et al.\ (2002) and M04 and is what 
one expects if the gas is in hydrostatic equilibrium and if its density profile 
traces that of the dark matter (e.g., Voit et al.\ 2002; Williams et al.\ 2004), 
an 
expectation that is supported by high resolution hydrodynamic simulations (e.g., 
Lewis et al.\ 2000; Voit et al.\ 2003; Ascasibar et al.\ 2003; Voit 2004).  As 
can 
be clearly seen, the slope of the dark matter profile, $\alpha$, is important in 
determining the resulting slope of the entropy profile, $\gamma$, within $0.1 
r_{halo}$ (which corresponds roughly to $r_{cool}$ for these clusters).  In 
particular, as the dark matter profile steepens the resulting entropy profile 
becomes more shallow, agreeing qualitatively and quantitatively with equation (3).

The remaining panels of Figure 1, which show the results for clusters that 
initially have steeper entropy profiles outside the entropy core, illustrate 
that the initial gas conditions of the clusters have almost no effect on the 
resulting steady-state entropy profile within $r_{cool}$.  Likewise, what 
happens in the interior of the cluster does not significantly influence gas 
outside of $r_{cool}$, as expected.

We have fitted the entropy profiles within $r_{cool}$ with powerlaws.  However, 
we find that the entropy profiles within $r_{cool}$ are not {\it exact} 
powerlaws and there is some ``wiggle'' room in the best fit logarithmic slope, 
depending on the range of radii over which the profiles are fitted.  For 
example, the best fit logarithmic slope over the range $0 \leq r \leq r_{cool}/2$ 
differs slightly from the best fit over the range $r_{cool}/2 \leq r \leq 
r_{cool}$.  We use these two radial intervals to roughly quantify the scatter in 
the best fit slope.  Figure 2 
presents a comparison between the analytic self-similar solution of B89 and 
the fits to the entropy profiles shown in Fig. 1.  The shaded region roughly 
reflects the uncertainty in the best fit powerlaw indices for the profiles 
predicted by M04's cooling model.

Reassuringly, excellent agreement between the self-similar solution and the 1-D 
cooling code is obtained.  Thus, the self-similar cooling wave model of 
B89 provides a physical basis for the powerlaw trends found by M04.  In 
addition, the agreement in Figure 2 gives us confidence in the reliability 
of the 1-D cooling model developed in M04.

\subsection{Realistic clusters}

Observed clusters and clusters formed in cosmological numerical simulations do 
not have pure powerlaw gravitational potentials.  Furthermore, the ICM contains
a significant quantity of metals and, consequently, cools not only through 
thermal bremsstrahlung but also through line emission.  Line emission 
has the effect of distorting the cooling function away from the powerlaw 
form that is characteristic of bremsstrahlung.  For these two reasons, the 
physical relevance of the results presented in \S 3.1 may be questioned.  Below, 
we investigate the extent to which the shape of the resulting entropy profile 
is affected by these assumptions.

\begin{figure*}
\centering
\includegraphics[width=13.0cm]{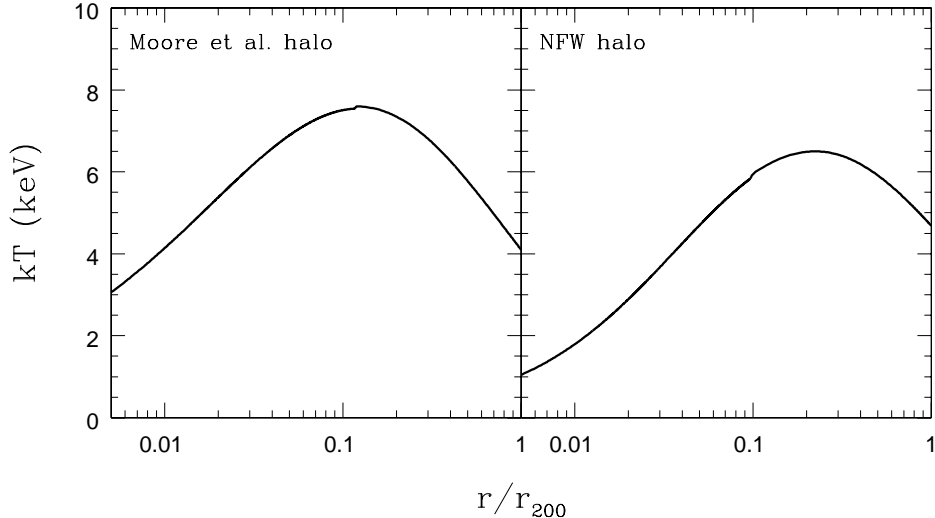}
\caption{
The final (steady-state) temperature distributions of the gas in the Moore et
al. and NFW halos.
}
\end{figure*}

Since we are introducing additional scales into the problem, it is important 
that we construct realistic cluster models.  We consider two different systems: 
one with a NFW dark matter profile and one with a Moore et al.\ dark matter 
profile.  Both systems have been chosen to have the same total gas and dark 
matter masses; specifically, $M_{gas}(r_{200}) \approx 1.5 \times 10^{14} 
M_{\odot}$ and $M_{dm}(r_{200}) = 10^{15} M_{\odot}$, where $r_{200} \approx 
2.06$ Mpc.  We use a typical cluster dark matter concentration of $c_{\rm 
NFW} \equiv r_{200}/r_s \approx 3.4$ for the NFW halo (e.g., Eke et al.\ 
1998; Bullock et al.\ 2001) and $c_{\rm Moore} \equiv r_{200}/r_{(-2)} = c_{\rm 
NFW}/0.630 \approx 5.4$ for the Moore et al.\ halo (see Keeton 2001).  The 
above implies a scale 
radius, $r_s$, of $\approx 600$ kpc.  As for the intracluster gas, we turn to 
the studies of Voit et al.\ (2003) and Voit (2004).  These authors found that 
the entropy profiles of a large sample of clusters generated with a numerical 
simulation of a $\Lambda$CDM cosmology including hydrodynamics (e.g., shock 
heating) but not radiative cooling are approximately self-similar over a 
wide range of masses (see Fig. 11 of Voit 2004).  The initial entropy 
distributions of our model clusters are assumed to be identical to 
Voit's best fit to his simulated clusters$^4$.  The initial gas density and 
temperature distributions are determined through the equation of hydrostatic 
equilibrium by applying the boundary condition that the total amount of gas 
within $r_{200}$ is equal to that specified above.  

\footnotetext[4]{At large radii, i.e., for $r > 0.1 r_{200}$, Voit 
(2004) reports a best fit entropy profile of $S(r) \propto r^{1.1}$.  At small 
radii, however, there is an apparent entropy core whose origin remains 
uncertain.  We have shrunk this core for computational convenience, since the 
model clusters reach steady state more quickly if they have small initial entropy 
cores.  However, this modification does not affect our results or conclusions 
since the entropy core is contained entirely within the cooling radius.  As 
discussed above, the resulting {\it steady-state} entropy profile within 
$r_{cool}$ depends only on the shapes of the gravitational potential and the 
cooling function (and not on the initial properties of the gas within 
$r_{cool}$). 
}

To compute the effects of cooling, we again make use 
of the model developed by M04.  In order to gauge the effects of line 
emission, we explore two different cooling functions: the pure thermal 
bremsstrahlung function implemented in \S 3.1 and a Raymond-Smith plasma with 
a metallicity set to $0.3 Z_{\odot}$.  As in the case of the powerlaw models, 
we neglect the dynamical effects of mass drop out and we run the cooling model 
until steady-state entropy profiles are achieved.  At steady state, we find 
that both clusters have similar global emission-weighted temperatures with 
$kT_{ew} \approx 5$ keV.

In Figure 3, we plot the initial and final entropy profiles of our model 
clusters.  In both panels, the dotted lines represent the initial entropy 
distributions, the thick solid lines represent the final distributions when 
cooled using a pure bremsstrahlung cooling function, and the thick dashed lines 
represent the final distributions when cooled using the $0.3 Z_{\odot}$ 
Raymond-Smith plasma cooling function.  The thin dashed lines indicate the 
slope predicted by B89's self-similar model assuming $\beta = 1/2$ and $\alpha 
= 1.5$ (left panel) and $\alpha = 1.0$ (right panel).  They have been 
arbitrarily normalized to cross the thick lines at the cooling radius.

First, consider the role of the cooling function.  For the Moore et al.\ halo, 
there is virtually no dependence on which cooling function we use.  The 
resulting entropy distribution for the NFW halo, however, is slightly shallower 
if we include line emission.  This difference can be understood as follows.  
The gravitational potential of the Moore et al.\ halo is steeper than that of 
the NFW halo.  Consequently, gas flowing into the center of the Moore 
et al.\ halo requires more thermal support to remain in hydrostatic 
equilibrium.  In Figure 4, 
we plot the final (steady-state) temperature distributions of the two model 
clusters.  Note that the central temperature of the Moore et al.\ halo is 
roughly 3 times larger than that of the NFW halo.  In Figure 5, we show the 
cooling function for a Raymond-Smith plasma with $Z = 0.3 Z_{\odot}$.  For 
temperatures of $kT \gtrsim 2$ keV, the cooling function is dominated by 
thermal bremsstrahlung and is well approximated by a powerlaw; $\Lambda(T) 
\propto T^{1/2}$.  This then explains why the entropy profile of the Moore et 
al. halo is unaffected by line emission: at any particular time there is 
virtually no gas below 2 keV (except at the exact center where gas rapidly 
cools below the X-ray emitting threshold of $\approx 10^5$ K).  The shallower 
NFW potential, however, permits some gas to cool below 2 keV before reaching 
the center of the cluster.  The thin dotted line in Fig. 5 shows the best fit 
powerlaw to the cooling function between $0.1$ keV $\leq kT \leq 2$ 
keV, which has an index of $\beta \approx -0.35$.  Using this value for 
$\beta$ in equation (2), we are able to account for the slight ($\sim 10$\%) 
deviation in the shape of the entropy profile within the central $0.01 
r_{200}$ ($\approx 20$ kpc) of the NFW halo.

The shape of the final entropy profile is more sensitive to the 
shape of the gravitational potential than it is to the shape of the cooling 
function.  Thus, we might expect self-similar solution to be a poor 
description of the final shape of entropy profiles in realistic dark matter 
halos.  However, the left hand panel of Fig. 3 
demonstrates that the shape of final entropy profile of the Moore et al.\ halo 
(thick lines) is virtually identical to that of a halo with a pure powerlaw 
profile of $\alpha = 1.5$ (thin dashed line).  Likewise, with the exception of 
the small deviation at the center due to line emission, the shape of the 
entropy profile in the NFW halo is essentially identical to that of a  
halo with a pure powerlaw profile of $\alpha = 1.0$ (see right hand panel of 
Fig. 3).  Recall that in the limit of $r \ll r_s$, the logarithmic slopes of 
the Moore et al.\ and NFW halos asymptote to $\alpha = 1.5$ and $1.0$, 
respectively.  Fig. 3 illustrates that what is relevant is the shape of 
the {\it local} gravitational potential (i.e., at $r \lesssim r_{cool}$), not 
the shape of the overall potential.  Thus, the self-similar solution 
provides an excellent description of steady-state cooling entropy profiles in 
realistic clusters {\it so long as the cooling radius is smaller than the 
cluster's scale radius}.  This condition should be met for most high mass 
clusters as the typical cooling radius of clusters is of order $\sim 100-200$ 
kpc (e.g., Peres et al.\ 1998), while the typical dark matter scale radius of 
massive clusters in high resolution cosmological simulations is of order $\sim 
400-700$ kpc (e.g., Eke et al.\ 1998; Bullock et al.\ 2001).

\begin{figure}
\centering
\includegraphics[width=8.4cm]{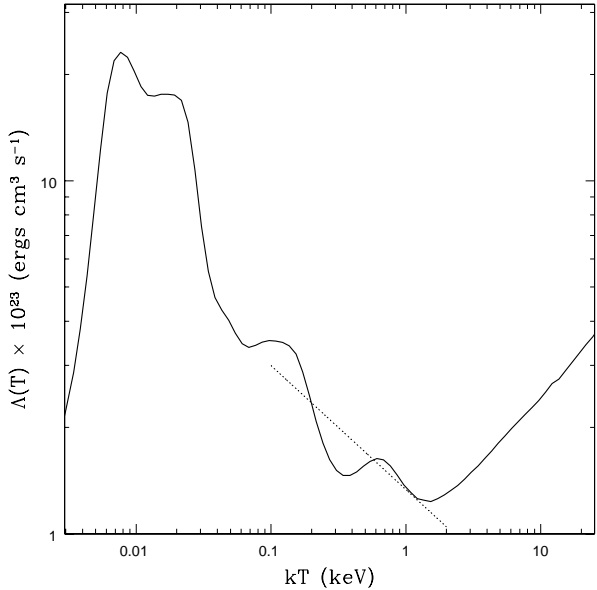}
\caption{
The cooling function for a Raymond-Smith plasma with $Z = 0.3 Z_{\odot}$.  The
dotted line represents the best fit powerlaw (with $\beta \approx -0.35$) to
the function over the range $0.1$ keV $\leq kT \leq 2.0$ keV.
}
\end{figure}

We conclude that the self-similar solution of B89 should provide a good 
description of the shapes of entropy profiles (for $r \lesssim r_{cool}$) of
massive clusters that cool via thermal bremsstrahlung and line emission and 
that have realistic dark matter profiles.  The reason for this is that 
self-similarity is only mildly violated for high mass clusters.  The deep 
gravitational potential wells of massive clusters ensure that most of the 
intracluster medium is quite hot ($kT \gtrsim 2$ keV), even within the cooling 
radius, and thus thermal bremsstrahlung (which is scale-free) dominates the 
X-ray emissivity.  The potential wells themselves have a characteristic scale 
radius but, for massive clusters, this radius is typically much larger than 
the cooling radius.  Therefore, the central cooling flow essentially ``feels'' 
only a pure powerlaw potential (as assumed by B89).

\section{Discussion}

Using the self-similar solution of B89, we have shown that radiative cooling 
and inflow lead to a characteristic entropy profile within the cooling radius 
that depends only on the shapes of the cooling function and the gravitational 
potential.  We have compared the self-similar solution to the cooling model of 
M04 and, reassuringly, find excellent agreement for clusters with powerlaw 
potentials and that cool via thermal bremsstrahlung.  Furthermore, we have 
demonstrated that the self-similar solution is also valid for more realistic 
dark matter potentials and cooling functions that include line emission, at 
least for massive clusters where bremsstrahlung dominates line emission and the 
typical dark matter scale radius is much larger than the cooling radius.  
This result explains the numerically-derived trends found in M04, which 
provide a good fit to a number of observed cooling flow clusters.  Kaiser 
\& Binney (2003) have also recently reported that cooling establishes a 
powerlaw trend between ICM gas mass and entropy.  This trend is also likely to 
be explained by the self-similar solution.  

The self-similar model of B89 implicitly assumes that there are no 
significant sources of non-gravitational heating present in the ICM.  Heating 
introduces an 
additional scale into the problem and, potentially, violates self-similarity.  It 
is clear, however, that some form of heating has (or is) occurred in real 
clusters in order to prevent the so-called cooling crisis (see Balogh et al.\ 
2001).  Additional evidence for heating comes from recent high resolution {\it 
Chandra} images of many cooling flow clusters that reveal 
buoyantly-rising bubbles of hot plasma (e.g., Heinz et al.\ 2002; Blanton et al.\ 
2003).  These bubbles were presumably blown by a central AGN and must be heating 
the ICM at some level.  Other sources of heating, such as thermal conduction 
(e.g., Medvedev \& Narayan 2001) and stirring due to the orbital motions of 
cluster galaxies (e.g., El-Zant et al.\ 2004), are also a possibility.  The 
relevant questions therefore are: (i) Is the level of heating sufficiently high 
to severely violate self-similarity?  (ii) If so, is the heating distributed or 
restricted to only the very center of the cluster?  (iii) In the event the 
heating is episodic (such as AGN heating), when did the last heating episode 
occur (i.e., has enough time passed in order to approximately re-establish a 
self-similar cooling flow)?  A number of relaxed clusters with published entropy 
profiles show 
evidence for large entropy cores (M04).  Clearly, such systems cannot be 
explained by the present self-similar cooling model, as these clusters were 
severely heated and the heating was distributed to large radii.  However, M04 
(see also Piffaretti et al. 2004) also found that several massive cooling flow 
clusters (e.g., A2029, PKS0745, Hydra A) have nearly pure powerlaw entropy 
profiles (except perhaps at very small radii, $r \lesssim 30$ kpc).  This likely 
indicates that the self-similar model provides a reasonably accurate description 
of the cooling gas in this particular subset of observed clusters. 

A test of the above hypothesis is to infer the slopes of the gravitational 
potentials of these clusters by using the slopes of their observed entropy 
profiles together with equation (3).  This may then be compared with the results 
obtained using the standard hydrostatic equilibrium method for inferring the 
gravitational potential of clusters.  Making use of the clusters with powerlaw 
entropy profiles from M04 (A2029, PKS0745, Hydra A), we find that the logarithmic 
slopes of their total matter density (dark matter and baryons) profiles are 
relatively steep, with $1.3 \lesssim \alpha \lesssim 2$.  This agrees quite well 
with the recent hydrostatic analysis of {\it Chandra} data of 10 relaxed 
cooling flow clusters by Arabadjis \& Bautz (2004).  Thus, for this small 
sample of clusters, the self-similar model appears to provide an apt description 
of the cooling gas in the centers of these clusters.  A more detailed comparison 
will soon be possible as the number of published cluster entropy profiles is 
rapidly increasing.

Quite independent of how well it describes observed clusters, the self-similar 
model also has a number of interesting {\it theoretical} uses.  We briefly 
discuss but two here.

1.) A simple method for calculating initial conditions of analytic model clusters 
with radiative cooling.  This could serve as a ``poor man's alternative'' to a 
model 
that explicitly takes into account the effects of radiative cooling and inflow 
on intracluster gas.  For example, a realistic set of initial conditions could 
be generated by using the results of non-radiative simulations (e.g., Lewis 
et al. 2000; Loken et al. 2002; Voit 2004) to describe the gas at large radii ($r 
> r_{cool}$) while 
using the self-similar solution of B89 to describe the properties of the gas 
within $r_{cool}$.  The normalization of the entropy profile within $r_{cool}$ 
(which is not specified by the self-similar solution) could be set by matching 
the non-radiative simulation results near $r_{cool}$.  One example of where 
such initial conditions might be useful is for models that explore the ability of 
various heating mechanisms to offset radiative losses of the ICM.  For 
example, a number of recent AGN heating simulations (e.g., Quilis, Bower, \& 
Balogh 2001; Ruszkowski \& Begelman 2002) have invoked initial conditions such as 
isothermality or temperature profiles derived from non-radiative simulation (as 
opposed to that expected for a cluster that is cooling radiatively) and this may 
have some effect on the estimates of the energetic requirements for the 
prevention of catastrophic cooling.  It would be interesting to see 
whether the estimates of the amount of required heating change significantly 
when more realistic initial conditions (such as those proposed above) are 
implemented.     

2.) A test of the reliability of cooling routines implemented in analytic 
models and hydrodynamic simulations.  Because the self-similar solution is a 
simple function, it can easily be used to test, for example, how well various 
formulations of smoothed particle hydrodynamics (SPH) or mesh-based techniques 
[such as adaptive mesh refinement (AMR)] treat the effects of cooling (see, 
e.g., Abadi, Bower, \& Navarro 2001).  We are currently undertaking such a 
study using a variety of popular analytic and hydrodynamic codes (Dalla 
Vecchia et al.\, in preparation).

\vskip 0.1in
\noindent 
I. G. M. is supported by a postgraduate scholarship from the Natural 
Sciences and Engineering Research Council of Canada (NSERC) and A. B. is 
supported by an NSERC Discovery Grant.  A. B. would also like to
acknowledge support from the Leverhulme Trust (UK) in the form of the
Leverhulme Visiting Professorship.


\begin{references}
\reference{}Abadi, M. G., Bower, R. G., \& Navarro, J. F. 2000, MNRAS, 314, 759
\reference{}Allen, S. W., Schmidt, R. W., \& Fabian A. C. 2001, MNRAS, 328, L37
\reference{}Arabadjis, J. S., \& Bautz, M. W. 2004, ApJ, submitted 
(astro-ph/0408362)
\reference{}Ascasibar, Y., Yepes, G., M\"{u}ller, V., \& Gottl\"{o}ber, S. 
2003, MNRAS, 346, 731
\reference{}Babul, A., Balogh, M. L., Lewis, G. F., \& Poole, G. B. 2002, 
MNRAS, 330, 329
\reference{}Balogh, M. L., Pearce, F. R., Bower, R. G., \& Kay, S. T. 2001,
MNRAS, 326, 1228
\reference{}Bertschinger, E. 1989, ApJ, 340, 666 (B89)
\reference{}Blanton, E. L., Sarazin, C. L., \& McNamara, B. R. 2003, ApJ, 585, 227
\reference{}Borgani, S., et al. 2004, MNRAS, 348, 1078
\reference{}Bullock, J. S., et al. 2001, ApJ, 550, 21
\reference{}David, L. P., et al. 2001, ApJ, 557, 546
\reference{}De Grandi, S., \& Molendi, S. 2002, ApJ, 567, 163
\reference{}Eke, V. R., Navarro, J. F., \& Frenk, C. S. 1998, ApJ, 503, 569
\reference{}El-Zant, A., Kim, W.-T., \& Kamionkowski, M. 2004, MNRAS, 354, 169
\reference{}Heinz, S., Choi, Y.-Y., Reynolds, C. S., \& Begelman, M. C. 2002, ApJ, 
569, L79
\reference{}Kaiser, C. R., \& Binney, J. 2003, MNRAS, 338, 837
\reference{}Keeton, C. R. 2001, ApJ, 561, 46
\reference{}Lewis, G. F., Babul, A., Katz, N., Quinn, T., Hernquist, L., \& 
Weinberg, D. H. 2000, ApJ, 536, 623
\reference{}Loken, C., Norman, M. L., Nelson, E., Burns, J., Bryan,
G. L., \& Motl, P. 2002, ApJ, 579, 571
\reference{}Matsushita, K., Belsole, E., Finoguenov, A., \& B\"{o}hringer, H. 
2002, A\&A, 386, 77
\reference{}McCarthy, I. G., Balogh, M. L., Babul, A., Poole, G. B., \& 
Horner, D. J. 2004, ApJ, 613, 811 (M04)
\reference{}Moore, B., Quinn, T., Governato, F., Stadel, J., \& Lake, G. 1999, 
MNRAS, 310, 1147
\reference{}Mushotzky, R., Figueroa-Feliciano, E., Loewenstein, M., \& 
Snowden, S. L. 2003, ApJ, submitted
\reference{}Narayan, R., \& Medvedev, M. V. 2001, ApJ, 562, L129
\reference{}Navarro, J. F., Frenk, C. S., \& White, S. D. M. 1997, ApJ, 490, 
493
\reference{}Oh, S. P., \& Benson, A. J. 2003, MNRAS, 342, 664
\reference{}Peres, C. B., Fabian, A. C., Edge, A. C., Allen, S. W.,
Johnstone, R. M., \& White, D. A. 1998, MNRAS, 298, 416
\reference{}Piffaretti, R., Jetzer, Ph., Kaastra, J. S., \& Tamura, T. 2004, 
A\&A, in press (astro-ph/0412233)
\reference{}Pratt, G. W., \& Arnaud, M. 2002, A\&A, 394, 375
\reference{}Quilis, V., Bower, R. G., \& Balogh, M. L. 2001, MNRAS, 328, 1091
\reference{}Ruszkowski, M., \& Begelman, M. C. 2002, ApJ, 581, 223
\reference{}Tozzi, P., \& Norman, C. 2001, ApJ, 546, 63
\reference{}Vikhlinin, A., Markevitch, M., Murray, S. S., Forman, W., \& Van 
Speybroeck, L. 2004, ApJ, submitted (astro-ph/0412306)
\reference{}Voit, G. M. 2004, Rev. Mod. Phys., in press (astro-ph/0410173) 
\reference{}Voit, G. M., Balogh, M. L., Bower, R. G., Lacey, C. G., \& Bryan,
G. L. 2003, ApJ, 593, 272
\reference{}Voit, G. M., Bryan, G. L., Balogh, M. L., \& Bower, R. G. 2002, 
ApJ, 576, 601
\reference{}Williams, L. R. L., Austin, C., Barnes, E., Babul, A., \& Dalcanton J. 
2004. To appear in the Proceedings of Science, published by SISSA; Conference: 
``Baryons in Dark Matter Haloes'', Novigrad, Croatia, 5-9 October 2004; editors: 
R.-J. Dettmar, U. Klein, P. Salucci (astro-ph/0412442)
\end{references}
\end{document}